\documentclass[lengthestimate,tighten,amssymb,prb,aps,twocolumn,floatfix,tightenlines,superscriptaddress]{revtex4}
\usepackage{graphicx,subfigure,color}
\newcommand{\ket}[1]{\left | #1 \right \rangle}

\begin{document}
\title{Manipulating multi-photon entanglement in waveguide quantum circuits}
\author{Jonathan C. F. Matthews}
\altaffiliation{These authors contributed to the work equally}
\affiliation{Centre for Quantum Photonics, H. H. Wills Physics Laboratory \& Department of Electrical and Electronic Engineering, University of Bristol, Merchant Venturers Building, Woodland Road, Bristol, BS8 1UB, UK}
\author{Alberto Politi}
\altaffiliation{These authors contributed to the work equally}
\affiliation{Centre for Quantum Photonics, H. H. Wills Physics Laboratory \& Department of Electrical and Electronic Engineering, University of Bristol, Merchant Venturers Building, Woodland Road, Bristol, BS8 1UB, UK}
\author{Andr\'{e} Stefanov}
\altaffiliation{Now at the Federal Office of Metrology METAS, Lindenweg 50, CH-3003 Bern-Wabern}
\affiliation{Centre for Quantum Photonics, H. H. Wills Physics Laboratory \& Department of Electrical and Electronic Engineering, University of Bristol, Merchant Venturers Building, Woodland Road, Bristol, BS8 1UB, UK}
\author{Jeremy L. O'Brien}
\email{Jeremy.OBrien@bristol.ac.uk}
\affiliation{Centre for Quantum Photonics, H. H. Wills Physics Laboratory \& Department of Electrical and Electronic Engineering, University of Bristol, Merchant Venturers Building, Woodland Road, Bristol, BS8 1UB, UK}
\date{\today}

\begin{abstract}
On-chip integrated photonic circuits are crucial to further progress towards quantum technologies and in the science of quantum optics. Here we report precise control of single photon states and multi-photon entanglement directly on-chip. We manipulate the state of path-encoded qubits using integrated optical phase control based on resistive elements, observing an interference contrast of $98.2\pm0.3\%$. We demonstrate integrated quantum metrology by observing interference fringes with 2- and 4-photon entangled states generated in a waveguide circuit, with respective interference contrasts of $97.2\pm0.4\%$ and $92\pm4\%$, sufficient to beat the standard quantum limit. Finally, we demonstrate a reconfigurable circuit that continuously and accurately tunes the degree of quantum interference, yielding a maximum visibility of $98.2\pm 0.9\%$.  These results open up adaptive and fully reconfigurable photonic quantum circuits not just for single photons, but for all quantum states of light.

\end{abstract}
\maketitle

\noindent Controlling quantum systems is not only a fundamental scientific endeavor, but promises profound new technologies \cite{nielsen, gi-rmp-74-145, gi-sci-306-1330}. Quantum photonics already provides enhanced communication security \cite{gi-rmp-74-145,matthews-note1}; has demonstrated increased precision by beating the standard quantum limit in metrology \cite{mi-nat-429-161, wa-nat-429-158, na-sci-316-726, hi-nat-450-393} and the diffraction limit in lithography \cite{bo-prl-85-2733, ka-oe-15-14249}; holds great promise for quantum computation \cite{kn-nat-409-46, ob-sci-318-1567}; and continues to advance fundamental quantum science. The recent demonstration of on-chip integrated waveguide quantum circuits \protect\cite{po-sci-320-646} is a key step towards these new technologies and for further progress in fundamental science applications.

Technologies based on harnessing quantum mechanical phenomena require methods to precisely prepare and control the state of quantum systems. Manipulation of a path-encoded qubit---a single photon in an arbitrary superposition of two optical paths, which is the natural encoding for waveguides \cite{po-sci-320-646}---requires control of the relative phase $\phi$ between the two optical paths and the amplitude in each path.

\begin{figure}[t!]
    \centering
    \includegraphics[width = \columnwidth]{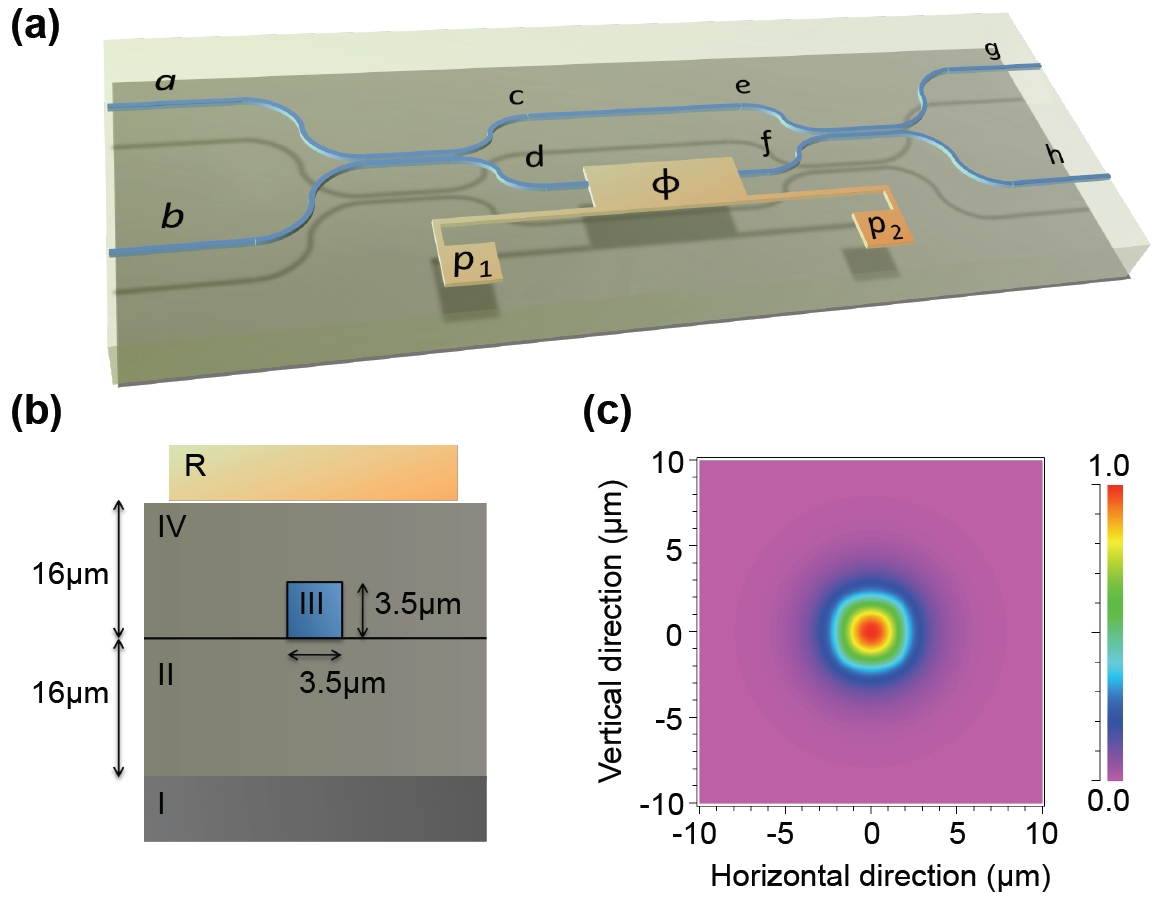}
\caption[]{\footnotesize{\textbf{Manipulating quantum states of light on a chip.} \textbf{a,} Schematic of a waveguide circuit with the relative optical phase $\phi$ controlled by applying a voltage $V$ across the contact pads $p_1$ and $p_2$ (not to scale). \textbf{b,} Illustration of the cross section of one waveguide located beneath a resistive heater. \textbf{c,} The simulated intensity profile of the guided single mode in the silica waveguides at a wavelength of 780nm. }}
\label{MZ}
\end{figure}

The integrated waveguide device shown schematically in Fig.~\ref{MZ}a applies the unitary operation $U_{MZ} = U_{DC} e^{i\phi \sigma_Z / 2}U_{DC}$: each 50\% splitting ratio (reflectivity $\eta=0.5$) directional coupler implements $U_{DC}$\cite{matthews-note6}; while control over the relative optical phase $\phi$ between the two optical paths implements the phase gate $e^{i \phi \sigma_z/2}$. A single photon input into mode \textit{a} is transformed into a superposition across modes \textit{c} and \textit{d}:
\begin{equation}
|1\rangle_a|0\rangle_b\rightarrow\frac{1}{\sqrt{2}}\left(|1\rangle_c|0\rangle_d+i|0\rangle_c|1\rangle_d\right)
\end{equation}
(a single photon input into mode $b$ is transformed into the same superposition but with a relative $\pi$ phase shift). The relative optical phase is then controlled by the parameter $\phi$, \emph{i.e.}
\begin{equation}
\frac{1}{\sqrt{2}}\left(|1\rangle_c|0\rangle_d+{i}|0\rangle_c|1\rangle_d\right)\rightarrow\frac{1}{\sqrt{2}}\left(|1\rangle_e|0\rangle_f+ie^{i\phi}|0\rangle_e|1\rangle_f\right)
\end{equation}
before the two modes are recombined at the second $\eta=0.5$ coupler.

The device shown in Fig.~\ref{MZ}a can also be used to manipulate the phase of multi-photon, entangled states of light. Two additional relative phase controllers before and after this device would enable arbitrary one-qubit unitary operations\cite{re-prl-73-58}, including state preparation and measurement \cite{matthews-note2}. By combining several such devices across $N$ waveguides, it is possible to realize
any arbitrary $N$-mode unitary operator \cite{re-prl-73-58}.

We begin by demonstrating a device which implements $U_{MZ}$, in which the phase shift $\phi$ is controlled by the voltage applied to a lithographically defined resistive heater. We then use this device to manipulate 1-, 2- and 4-photon entangled states relevant to quantum metrology. Finally, we demonstrate how such a device can be used to realize a reconfigurable photonic quantum circuit.

\begin{large}
\vspace{12pt}{\noindent\textbf{Results}}\\
\end{large}
\noindent\textbf{Voltage-controlled phase shift.}
Waveguide devices, as illustrated in cross-section in Fig.~\ref{MZ}b, were fabricated on a 4" silicon wafer (material \textit{I}), onto which a 16 $\mu$m layer of thermally grown undoped silica was deposited as a buffer to form the lower cladding of the waveguides (\textit{II}). A 3.5 $\mu$m layer of silica doped with germanium and boron oxides was then deposited by flame hydrolysis; the material of this layer constitutes the core of the stucture and was patterned into 3.5 $\mu$m wide waveguides via standard optical lithographic techniques (\textit{III}). The 16 $\mu$m upper cladding (\textit{IV}) is  phosphorus and boron doped silica with a refractive index matched to that of the buffer. Simulations indicated single mode operation at 780 nm, as shown in Fig.~\ref{MZ}c. A final metal layer was lithographically patterned on the top of the devices to form  resistive elements ($R$) and the metal connections and contact pads ($p_1$ and $p_2$) shown in Fig.~\ref{MZ}a.

When a voltage is applied between $p_1$ and $p_2$, the current in $R$ generates heat which dissipates into the device and locally raises the temperature $T$ of the core and cladding material of the waveguide section directly below. To first approximation, the change in refractive index $n$ of silica is given\cite{kenichi} by $dn/dT = 10^{-5}/$K (independent of compositional variation) which in turn alters the mode group index of the light confined in the waveguide beneath \textit{R}. The devices were designed to enable a continuously variable phase shift $\phi\in[-\pi/2,\pi/2]$ and operate at room temperature. A consequence of the miniature and monolithic structure of the chip is that no strict global temperature control of the device is required for stability (see supplementary information).

\begin{figure}
\begin{center}
\includegraphics[width=8.5cm]{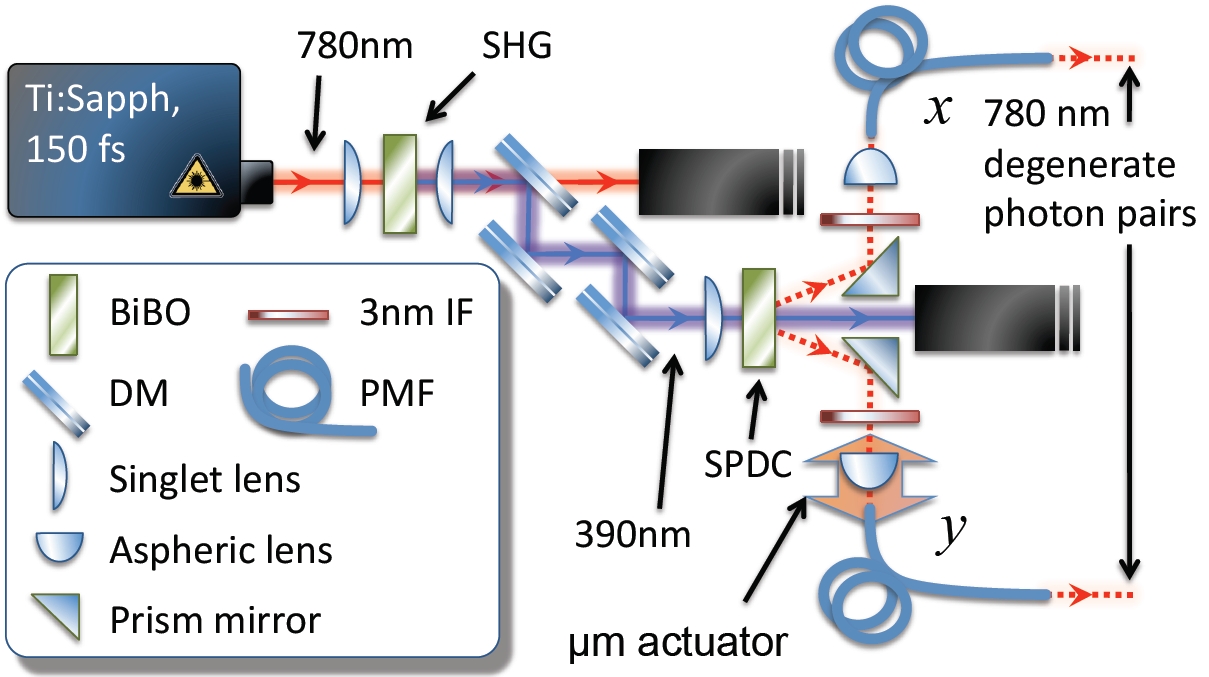}
\caption{\footnotesize{\textbf{Multi photon state preparation.}\\ Pulsed, coherent 390 nm light pumps a Type-I nonlinear Bismuth Borate $\textrm{BiB}_3\textrm{O}_6$ crystal for spontaneous parametric down conversion. Depending on the average pump power, we produce 2- and 4-photon states of 780 nm degenerate photons in two paths. See Methods for further details.}}
\label{figS4}
\end{center}
\end{figure}

\begin{figure}[t]
     \centering
      \includegraphics[width=8cm]{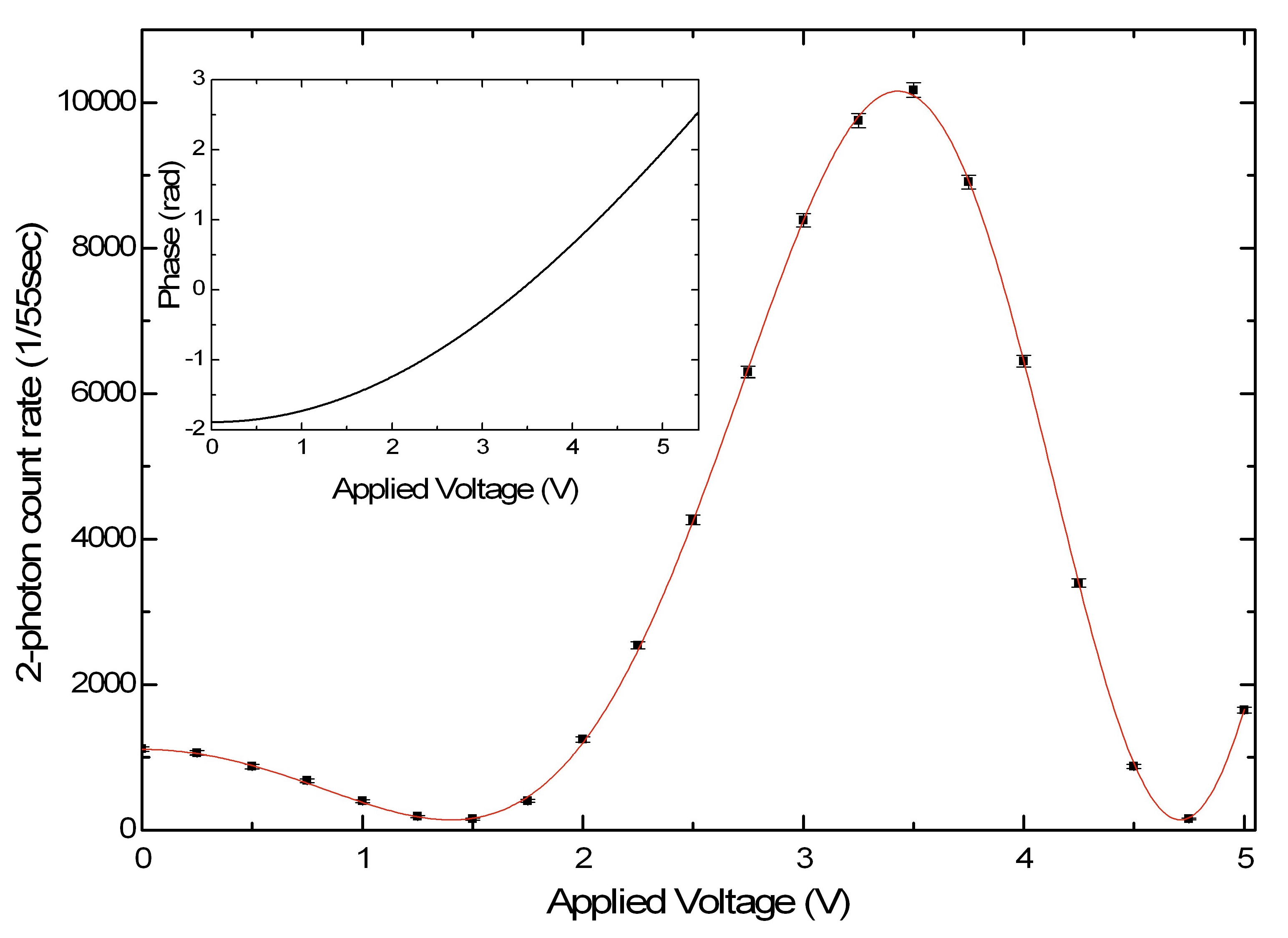}
     \caption{\textbf{Calibration of voltage-controlled phase shift.}\\ \textbf{Main panel,} The {2}-photon interference pattern generated from simultaneous detection of a single photon at both outputs $g$ and $h$ as the voltage applied across the device was varied between 0 and 5V. Error bars are given by Poissonian statistics.
     \textbf{Inset,} Plot of  the phase-voltage relationship determined from this calibration.}
     \label{voltage}
\end{figure}

The voltage-controlled phase inside the waveguide circuit, shown schematically in Fig.~\ref{MZ}a, is determined by a nonlinear relation $\phi\left(V\right)$, which we calibrated using a 2-photon interference effect (see supplementary information):
ideally, the maximally path entangled state of two photons
\begin{equation}
\frac{1}{\sqrt{2}}\left(\ket{2}_c\ket{0}_d+\ket{0}_c\ket{2}_d\right)
\label{2noon}
\end{equation}
 is generated inside the device\cite{ou-pra-42-2957,ra-prl-65-1348,ku-qso-10-493, fo-prl-82-2868, ed-prl-89-213601, ei-prl-94-090502, re-prl-98-223601} on inputting the state $\ket{1}_a\ket{1}_b$, which we produced using the setup shown in Fig. \ref{figS4}. After the phase shift this entangled state is transformed to $\frac{1}{\sqrt{2}}(\ket{2}_e\ket{0}_f+ e^{i2\phi}\ket{0}_e\ket{2}_f)$. Fig.~\ref{voltage} shows the results of this calibration in which the rate of simultaneous detection of two photons at outputs $g$ and $h$ is plotted as a function of the applied voltage $V$ across $p_1$ and $p_2$. The phase voltage relationship was verified to be a polynomial function of the form:
\begin{eqnarray}
\phi\left(V\right) = \alpha + \beta V^2 + \gamma V^3 + \delta V^4,
\label{PhasevsVoltage}
\end{eqnarray}
where the parameters were found by means of best-fit (see supplementary information); the resulting relationship is plotted in the inset of Fig.~\ref{voltage}. In comparison to simply using 1-photon ``classical" interference, this ``quantum calibration'' harnesses the reduced de Broglie wavelength \cite{mi-nat-429-161, na-sci-316-726,wa-nat-429-158} of 2-photon interference \cite{ou-pra-42-2957,ra-prl-65-1348,ku-qso-10-493, fo-prl-82-2868, ed-prl-89-213601, ei-prl-94-090502, re-prl-98-223601} to wider sample the pattern of phase dependent interference, thereby giving greater precision in the $\phi(V)$ calibration \cite{matthews-note4}. The phase shift was found to be stable on the several hours timescale (see supplementary information).

\begin{figure}[t]
     \centering
      \includegraphics[width=8cm]{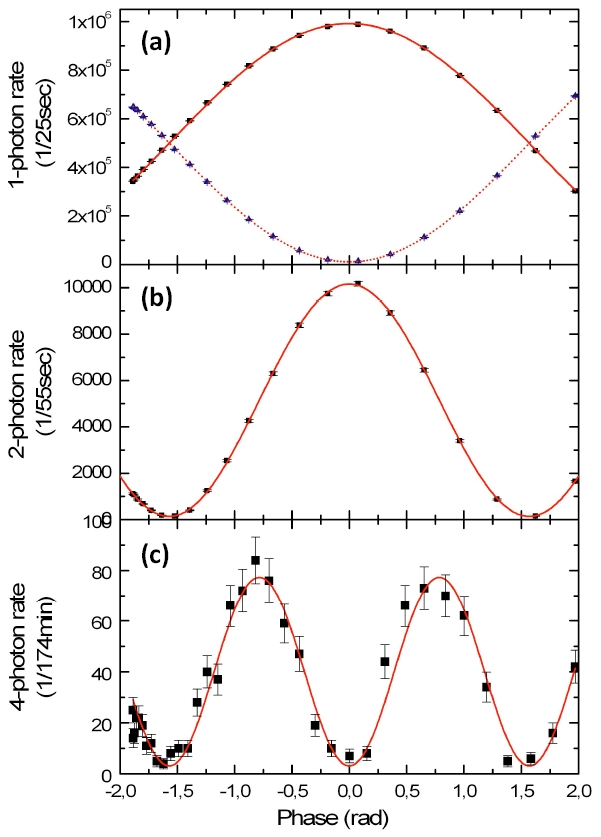}
     \caption{\textbf{Integrated quantum metrology.}\\ \textbf{(a)} {1-}photon count rates {at} the outputs $g$ {(blue triangle data points, dotted fit)} and $h$ {(black square data points, solid fit)} as the phase $\phi\left(V\right)$ is varied {on inputting the 1-photon state $\ket{1}_a\ket{0}_b$}. \textbf{(b)} {2}-photon coincidental detection rate between the outputs $g$ and $h$ when inputting the {2-}photon state $\ket{1}_a\ket{1}_b$ and varying the phase $\phi\left(V\right)$. \textbf{(c)} {4}-photon coincidental detection rate of the output state $\ket{3}_g\ket{1}_h$ when inputting the {4}-photon state $\ket{2}_a\ket{2}_b$. Error bars are given by Poissonian statistics.}
     \label{fringeplots}
\end{figure}

\vspace{10pt}\noindent\textbf{Multi-photon entangled state manipulation.}
Having obtained $\phi(V)$, we were able to analyze the sinusoidal interference pattern arising from single photon detections at  outputs $g$ and $h$ when launching single photons into input \textit{a} and controlling $\phi(V)$. Ideally the probability of detecting photons varies as
\begin{equation}
P_{g}  =1-P_{h} = \frac{1}{2}\left[1 {-}\cos{\left(\phi\right)}\right],
\end{equation}
yielding sinusoidal interference fringes with a period of $2\pi$. The observed fringes (Fig.~\ref{fringeplots}a) show a high contrast \cite{matthews-note5} of $C=0.982\pm0.003$. From this contrast, and assuming no mixture or complex phase is introduced, it is possible to calculate the average fidelity $F$ between the measured and ideal output state $U_{MZ}\ket{0}=\cos\left(\phi/2\right)\ket{0}+{i}\sin\left(\phi/2\right)\ket{1}$. Averaging over the range $\phi\in\left[-\pi/2,\pi/2\right]$ we find $\overline{F}=0.99984\pm 0.00004$.

{
These devices also enable us to manipulate and analyze multi-photon entangled states: \emph{eg.} the state (\ref{2noon}) should ideally be} transformed according to $\frac{1}{\sqrt{2}}(\ket{2}_e\ket{0}_f {+} e^{2i\phi}\ket{0}_e\ket{2}_f)$. To confirm the correct on-chip control of this entangled state, simultaneous detection of a single photon at each output $g$ and $h$ {was recorded} as a function of $\phi$; this {ideally} yields a ``$\lambda/2$" interference fringe described by
\begin{equation}
P_{g,h} = \frac{1}{2}\left(1{+}\cos{2\phi}\right)
\end{equation}
with period $\pi$---half the period of the {1}-photon interference fringes.
{The 2}-photon interference fringe shown in Fig.~\ref{fringeplots}b plots the {measured} simultaneous detection rate as a function of $\phi$. The contrast is $C=0.972\pm0.004$, which is greater than the threshold $C_{th}=1/\sqrt{2}$ required to beat the standard quantum limit\cite{ok-njp-10-073033}, as described below.
{Note that although a 2-photon interference fringe was used to calibrate the phase shift, this calibration is not required to claim a $\lambda/2$ interference fringe; this is simply confirmed by comparison with the 1-photon fringe, which can be done even without calibrating the phase.}

The interference fringe shown in Fig.~\ref{fringeplots}b arises from the {2}-photon maximally path entangled state that is an equal superposition of $N$ photons in one mode and $N$ photons in another mode: $\ket{N}\ket{0}+\ket{0}\ket{N}$ \cite{le-jmo-49-2325}. Such a state evolves under a $\phi$ phase shift in the second mode to $\ket{N}\ket{0}+e^{iN\phi}\ket{0}\ket{N}$ and can in principle be used to estimate an unknown phase $\phi$ with a sensitivity $\Delta\phi=1/N$, better than the standard quantum limit $\Delta\phi=1/\sqrt{N}$ --- the limit attainable with classical schemes. By inputting the {4}-photon state  $\ket{2}_a\ket{2}_b$ non-classical interference at the first directional coupler ideally produces the state \cite{st-pra-65-033820, na-sci-316-726, ok-njp-10-073033}
\begin{eqnarray}
\sqrt{\frac{3}{4}}\left(\ket{4}_c\ket{0}_d + \ket{0}_c\ket{4}_d\right)/\sqrt{2} {+} \frac{1}{\sqrt{4}}\ket{2}_c\ket{2}_d.
\label{4-phstateMZ}
\end{eqnarray}
At the second directional coupler, quantum interference means that only the $\ket{4}_c\ket{0}_d + \ket{0}_c\ket{4}_d$ part of this state gives rise to $\ket{3}_e\ket{1}_f$ and $\ket{1}_e\ket{3}_f$ in the output state of the interferometer. By varying the phase $\phi$ in the interferometer, the probability of detecting either of the states $\ket{3}_e\ket{1}_f$ or $\ket{1}_e\ket{3}_f$ is given by
\begin{equation}
P_{3g,h}=P_{g,3h}=\frac{3}{8}\left(1-\cos{4\phi}\right)
\end{equation}
and yields a ``$\lambda/4$'' interference fringe with period $\pi/2$. We measured the {4}-photon interference fringe shown in Fig.~\ref{fringeplots}c, which plots the rate of simultaneous detection of four photons corresponding to the state $\ket{3}_g\ket{1}_h$ (by cascading three detectors using $1\times2$ fibre-beam splitters at the output $g$) against the phase $\phi$. The contrast of this {4}-photon interference is $C=0.92\pm 0.04$, which is greater than the threshold to beat the standard quantum limit \cite{ok-njp-10-073033}.

\begin{figure} 	
    \centering
    \includegraphics[width = 8.5cm]{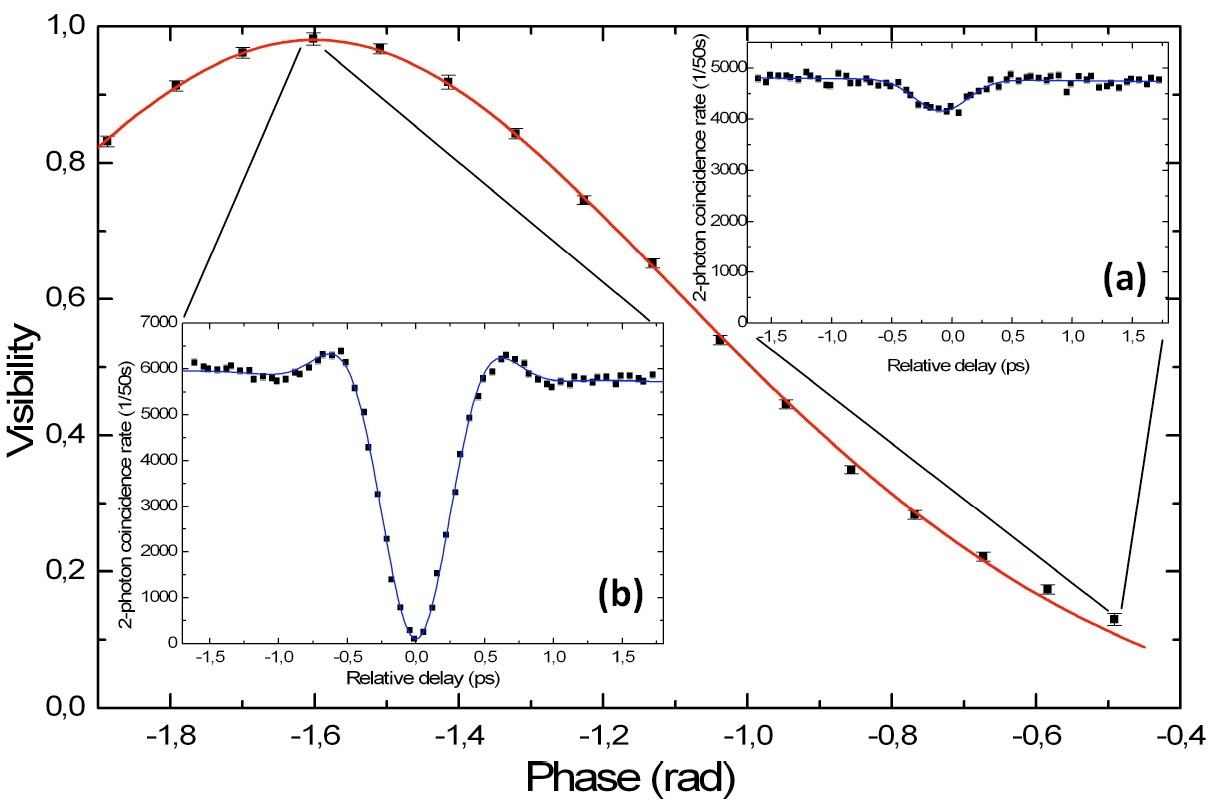}
\caption[]{\footnotesize{\textbf{A reconfigurable quantum circuit}\\ \textbf{Main panel,} Visibility of the Hong-Ou-Mandel experiment performed using the integrated MZ interferometer as a continuously variable beam splitter with effective reflectivity $\eta = \sin^2\left(\phi/2\right)$. The solid line is a theoretical fit that includes a small phase-offset and a small amount of mode-mismatch as the only two free parameters that modify Eqs. \ref{eta} and \ref{visibility},. Error bars for the main panel are given by confidence intervals on the best-fit parameter. \textbf{Inset a,} High visibility {2}-photon interference. \textbf{Inset b,} Low visibility {2}-photon interference. Both inset plots are displayed as a plot of {2}-photon rate versus the relative optical delay between the interfering photons and fitted with a function that takes into account the non gaussian shape of the interference filter used in the experiment. Error bars for each inset are given by Poissonian statistics.}}
\label{VariableBS}
\end{figure}

\vspace{10pt}\noindent\textbf{Reconfigurable quantum circuits.}
Quantum interference of photons \cite{ho-prl-59-2044} at a directional coupler (or beamplitter) lies at the heart of the multi-photon interference fringes shown in Fig.~\ref{fringeplots} and is the crucial underlying physical process in linear optical networks for quantum information science. The reflectivity $\eta$ of a coupler determines the degree of quantum interference, thereby making $\eta$ the critical parameter for quantum operation. The directional couplers in the device shown schematically in Fig. \ref{MZ} were lithographically set to $\eta=\frac{1}{2}$.  More general photonic circuits, including optical entangling logic gates \cite{kn-nat-409-46, ob-nat-426-264, po-sci-320-646, la-nphys-5-134}, are composed of a number of different reflectivity couplers, while adaptive schemes whose function depends on the input state, such as Fock state filters \cite{sa-prl-96-083601, re-prl-98-203602, ok-sci-323-483}, make use of devices equivalent to a single coupler with variable $\eta$. Reconfigurable photonic circuits, including routing of photons, can be realized by combining such variable $\eta$ devices. By controlling the phase $\phi$ within our devices, we implement the unitary operation
\begin{eqnarray}
U_{MZ} \stackrel{.}{=} \left(
\begin{array}{l r}
\sin \left(\phi/2\right) & \cos \left(\phi/2\right) \\
\cos \left(\phi/2\right) & {-}\sin \left(\phi/2\right)
\end{array}
\right),
\end{eqnarray}
acting on the two input waveguides\cite{re-prl-73-58}. This operation is equivalent to a single coupler with variable reflectivity
\begin{equation}
\label{eta}
\eta = \sin^2{\frac{\phi}{2}}.
\end{equation}

We performed multiple quantum interference experiments\cite{ho-prl-59-2044} in which two photons were launched into inputs $a$ and $b$ of the device. While scanning through the relative arrival time with an off-chip optical delay, we measured the rate of simultaneous detection of a single photon at both outputs $g$ and $h$. Each experiment resulted in a quantum interference ``dip" in this rate of simultaneous photon detection, centred around zero delay (\emph{eg.} see insets to Fig.~\ref{VariableBS}). The depth of such a dip indicates the degree of quantum interference, which can be quantified by the visibility $V=(N_{max}-N_{min})/N_{max}$. Ideally
\begin{equation}
\label{visibility}
V_{ideal}=\frac{2\eta(1-\eta)}{1 - 2 \eta + 2 \eta^2}.
\end{equation}
The main panel in Fig.~\ref{VariableBS} plots the quantum interference visibility observed for different values of $\phi$ and hence $\eta$. The insets of Fig.~\ref{VariableBS} show two examples of the raw data used to generate this curve: ({a}) $\phi = -0.49 \pm 0.01$ radians, $V=0.129\pm 0.009$; ({b}) $\phi = -1.602 \pm 0.01$ radians, $V=0.982\pm 0.009$.
The average relative visibility $V_{rel}=V/V_{ideal}$ for all of the data in Fig.~\ref{VariableBS} is $\overline{V_{rel}}=0.980\pm 0.003$

\begin{large}
\vspace{12pt}{\noindent\textbf{Discussion}}\\
\end{large}
Integrated optics has been developed primarily by the telecommunications industry for devices that allow high-speed information transmission, including optical switches, wavelength division multiplexers, and modulators.
Quantum optics appears destined to benefit from existing integrated optics technologies, as well as drive new developments for its own needs. The reconfigurable quantum circuit we demonstrate could be used as the fundamental element to build a large-scale circuit capable of implementing any unitary operation on many waveguides. A thermal-based 32$\times$32 waveguide switch has been demonstrated\cite{32}. Implementing an arbitrary unitary on this number of modes would require a comparable number of resistive elements. This is well beyond anything conceivable with bulk optics.  The ms timescales available with thermal switching is suitable for reconfigurable circuits, for state preparation, quantum measurement, quantum metrology\cite{gi-sci-306-1330}, and perhaps even full-scale quantum computing\cite{ke-prl-99-130501}. Other applications demanding fast switching, such as adaptive circuits for quantum control and feedforward, will require sub ns switching, which is possible using electro-optic materials such as LiNbO$_3$, used to make modulators operating at 10's of GHz\cite{wo-jtsqe-6-69}.

In addition to the demonstrations presented here, these devices may be used for other quantum states of light, for the fundamental sciences of quantum optics \cite{lo-sci-322-563, fu-sci-282-706, pa-sci-317-1890} and quantum information \cite{br-rmp-77-513, ou-sci-312-83,me-prl-101-130501}. In particular, phase control will be particularly important for homodyne detection required for phase estimation \cite{pe-prl-99-223602} and adaptive measurements \cite{be-pra-73-063824}  with squeezed states of light. Our results point towards adaptive and arbitrarily reconfigurable quantum networks capable of generating, manipulating and characterizing multi-photon states of light with near-unit fidelity. Possible future applications span all of quantum information science from metrology to information processing.

\vspace{20pt}
\begin{large}
{\noindent\textbf{Methods}}\\
\end{large}
\begin{footnotesize}
{
\noindent\textbf{Devices.}
The bend radius of curves in the directional couplers in the waveguide circuit are 15 mm at the tightest curvature, the effective interaction length of each directional coupler is $2.95$ mm, while each path within the interferometer is $9.7$ mm (defined from the end of the first directional coupler to the beginning of the second directional coupler) and the maximum optical path difference with the maximum voltage we apply is $\sim\lambda/2$ (i.e.$\sim390$ nm). The physical length of the chip from input facet to output facet is 26 mm.
}\vspace{10pt}\\
{\noindent\textbf{Multi-photon generation.} The experiments reported were conducted using degenerate single photon pairs at a wavelength of 780 nm produced via spontaneous parametric down conversion (SPDC). The nonlinear crystal used is a Type-I phase matched Bismuth Borate $\textrm{BiB}_3\textrm{O}_6$ (BiBO) pumped by a 390 nm 150 fs pulsed laser focused to a waist of $\omega_0\approx 40$ $\mu\textrm{m}$. The 390 nm pump was prepared using a further BiBO crystal, phase matched for second harmonic generation (SHG) to double the frequency of a 780 nm mode-locked Ti:Sapphire laser focused to a waist of $\omega_0\approx 40$ $\mu\textrm{m}$; four successive dichroic mirrors (DM) are used to purify the pump beam spectrally. Degenerate photon pairs are created by the SPDC crystal and pass through 3 nm interference filters (IF) which filter each photon to a coherence length of $l_c=\lambda^2/\Delta \lambda\approx 200$ $\mu \textrm{m}$. The photons are collected into two single mode polarization maintaining fibers (PMFs) coupled to two diametrically opposite points $x$ and $y$ on the SPDC cone. In the case of low average pump power, the state $\ket{1}_x\ket{1}_y$
is produced with a rate of 100 $\textrm{s}^{-1}$. On increasing the average pump power, the multi-photon production rate from the down-conversion process  is no longer negligible such that it is possible to produce two degenerate pairs of photons in the state $\ket{2}_x\ket{2}_y$.
}\vspace{10pt}\\
{
\noindent\textbf{Coupling to devices.}
The photons coupled into PMF were launched into the chip and collected at the outputs using two arrays of 8 PMF, with 250 $\mu$m spacing, to match that of the waveguides. The photons were detected using fiber coupled single photon counting modules (SPCMs). The PMF arrays and chip were  directly buttcoupled, with index matching fluid, to obtain an overall coupling efficiency of $\sim$60\% through the device (input plus output insertion losses $\sim$40\%).
}\\

\end{footnotesize}

\bibliographystyle{unsrtnat}

\begin{thebibliography}{1}
\vspace{12pt}
\begin{large}
{\noindent\textbf{References}}\\
\end{large}

\bibitem{nielsen} Nielsen, M. A. and Chuang, I. L. \textit{Quantum Computation and Quantum Information.} (Cambridge University Press, 2000).
\bibitem{gi-rmp-74-145} Gisin, N., Ribordy, G., Tittel, W. \& Zbinden, Quantum cryptography. \textit{Rev. Mod. Phys.} \textbf{74}, 145-195 (2002).
\bibitem{gi-sci-306-1330} Giovannetti, V., Lloyd, S., \& Maccone, L. Quantum-Enhanced Measurements: Beating the Standard Quantum Limit. \textit{Science} \textbf{306}, 1330-1336 (2004).
\bibitem{matthews-note1} See the recent demonstration of the SECOQC network in Vienna: www.secoqc.net (2008).
\bibitem{mi-nat-429-161} Mitchell, M. W., Lundeen, J. S. \& Steinberg, A. M. Super-resolving phase measurements with a multiphoton entangled state. \textit{Nature} \textbf{429}, 161-164 (2004).
\bibitem{wa-nat-429-158} Walther, P., Pan, J. W., Aspelmeyer, M., Ursin, R., Gasparoni, S. \& Zeilinger, A.  De Broglie wavelength of a non-local four-photon state. \textit{Nature} \textbf{429}, 158-161 (2004).
\bibitem{na-sci-316-726} Nagata, T., Okamoto, R. O'Brien, J. L., Sasaki, K. \& Takeuchi, S. Beating the Standard Quantum Limit with Four-Entangled Photons. \textit{Science} \textbf{316}, 726-729 (2007).
\bibitem{hi-nat-450-393} Higgins, B. L., Berry, D. W., Bartlett, S. D., Wiseman, H. M. \& Pryde, G. J. Entanglement-free Heisenberg-limited phase estimation. \textit{Nature} \textbf{450}, 393-396 (2007).
\bibitem{bo-prl-85-2733} Boto, A. N., Kok, P., Abrams, D. S., Braunstein, S. L., Williams, C. P. \& Dowling, J. P. Quantum  Interferometric Optical Lithography: Exploiting Entanglement to Beat the Diffraction Limit. \textit{Phys. Rev. Lett.} \textbf{85}, 2733-2736 (2000).
\bibitem{ka-oe-15-14249} Kawabe, Y., Fujiwara, H., Okamoto, R., Sasaki, K \& Takeuchi, S. Quantum interference fringes beating the diffraction limit. \textit{Opt. Express} \textbf{15}, 14244-14250 (2007).
\bibitem{kn-nat-409-46} Knill, E., Laflamme, R. \& Milburn, G. J. A scheme for efficient quantum computation with linear optics. \textit{Nature} \textbf{409}, 46-52 (2001).
\bibitem{ob-sci-318-1567} O'Brien, J. L. Optical quantum computing. \textit{Science} \textbf{318}, 1567-1570 (2007).
\bibitem{po-sci-320-646} Politi, A., Cryan, M. J., Rarity, J. G., Yu, S. \& O'Brien, J. L. Silica-on-Silicon Waveguide Quantum Circuits. \textit{Science} \textbf{320}, 646-649 (2008).

\bibitem{matthews-note6} {$U_{dc}=e^{i\pi/2} e^{-i\pi\sigma_{Z}/4} H e^{-i\pi\sigma_{Z}/4}$ 
transforms the logical qubit states according to $U_{DC}\ket{0} = \frac{1}{\sqrt{2}}\left(\ket{0}+i\ket{1}\right)$ and $U_{DC}\ket{1} = \frac{1}{\sqrt{2}}\left(i\ket{0}+\ket{1}\right)$; $\left\{\sigma_{X},\sigma_{Y},\sigma_{Z}\right\}$ are the single qubit Pauli operators.}

\bibitem{re-prl-73-58} Reck, M., Zeilinger, A., Bernstein, H. J. \& Bertani, P. Experimental realization of any discrete unitary operator. \textit{Phys. Rev. Lett.} \textbf{73}, 58-61 (1994).

\bibitem{matthews-note2} {First note the relations $\sigma_{X} e^{i\phi \sigma_{Z} /2} =e^{-i\phi \sigma_{Z} /2}\sigma_{X}$ and  $U_{MZ}=i e^{i\phi \sigma_{Y} /2} \sigma_{X} $.} {For some real $\phi_{1},\phi_{2},\phi_{3}$, a}rbitrary qubit operations can be decomposed as {$U_{arb} = e^{i\phi_3 \sigma_Z / 2}e^{i\phi_2 \sigma_Y / 2}e^{i\phi_1 \sigma_Z / 2}$}; arbitrary qubit preparation from the logical basis is applied by {$U_{prep}=e^{i\phi_3 \sigma_{Z}/2 }e^{i\phi_2 \sigma_{Y} / 2}$}; the inverse (or time-reversed) operation $U^{\dagger}_{prep}$ provides arbitrary projective measurement\cite{nielsen}.

\bibitem{kenichi} Kenichi, I. \& Yokubun, Y. \textit{Encyclopedic Handbook of Integrated Optics.} (CRC Press, 2006).
\bibitem{ou-pra-42-2957}  Ou, Z. Y., Zou, X. Y., Wang, L. J. \& Mandel, L. Experiment on nonclassical fourth-order interference. \textit{Phys. Rev. A} \textbf{42}, 2957-2965 (1990).
\bibitem{ra-prl-65-1348} Rarity, J. G., Tapster, P. R., Jakeman, E., Larchuk, T., Campos, R. A., Teich, M. C. \& Saleh, B. E. A. Two-photon interference in a Mach-Zehnder interferometer. \textit{Phys. Rev. Lett.} \textbf{65}, 1348-1351 (1990).
\bibitem{ku-qso-10-493} Kuzmich, A. \& Mandel, L. Sub-shot-noise interferometric measurements with two-photon states. \textit{Quant. Semiclass. Opt.} \textbf{10}, 493-500 (1998).
\bibitem{fo-prl-82-2868} Fonseca, E. J. S., Monken, C. H. \& P\'adua, S. Measurement of the de Broglie Wavelength of a Multiphoton Wave Packet. \textit{Phys. Rev. Lett} \textbf{82}, 2868-2871 (1999).
\bibitem{ed-prl-89-213601} Edamatsu, K., Shimizu, R. \& Itoh, T. Measurement of the Photonic de Broglie Wavelength of Entangled Photon Pairs Generated by Spontaneous Parametric Down-Conversion. \textit{Phys. Rev. Lett.} \textbf{89}, 213601 (2002).
\bibitem{ei-prl-94-090502} Eisenberg, H. S., Hodelin, J. F., Khoury, G. \& Bouwmeester, D. Multiphoton Path Entanglement by Nonlocal Bunching. \textit{Phys. Rev. Lett.} \textbf{94}, 090502 (2005).
\bibitem{re-prl-98-223601} Resch, K. J., Pregnell, K. L., Prevedel, R., Gilchrist, A., Pryde, G. J., O'Brien, J. L. \& White, A. G. Time-Reversal and Super-Resolving Phase Measurements. \textit{Phys. Rev. Lett.} \textbf{98}, 223601 (2007).
\bibitem{matthews-note4} We were limited in the range of $\phi$ we could implement by the maximum voltage that can be applied across the electrodes.
\bibitem{matthews-note5} The contrast is defined as $C=(N_{max}-N_{min})/(N_{max}+N_{min})$.
\bibitem{ok-njp-10-073033} Okamoto, R., Hofmann, H. F., Nagata, T., O'Brien, J. L., Sasaki, K. \& Takeuchi, S. Beating the standard quantum limit: phase super-sensitivity of N-photon interferometers. \textit{New J. Phys.} \textbf{10} 073033 (2008).
\bibitem{le-jmo-49-2325} Lee, H., Kok, P. \& Dowling, J. P. A quantum Rosetta stone for interferometry. \textit{J. Mod. Opt.} \textbf{49}, 2325-2338 (2002).
\bibitem{st-pra-65-033820} Steuernagel, O. De Broglie wavelength reduction for a multiphoton wave packet. \textit{Phys. Rev. A} \textbf{65}, 033820 (2002).
\bibitem{ho-prl-59-2044} Hong, C. K., Ou, Z. Y. \& Mandel, L. Measurement of subpicosecond time intervals between two photons by interference. \textit{Phys. Rev. Lett.} \textbf{59}, 2044-2046 (1987).
\bibitem{ob-nat-426-264} O'Brien, J. L., Pryde, G. J., White, A. G., Ralph, T. C. \& Branning, D. Demonstration of an all-optical quantum controlled-\textsc{NOT} gate. \textit{Nature} \textbf{426}, 264-267 (2003).

\bibitem{la-nphys-5-134} Lanyon, B. P., Barbieri, M., Almeida, M. P., Jennewein, T., Ralph, T. C., Resch, K.~J., Pryde, G.~J., O'Brien, J.~L., Gilchrist, A. \&  White, A.~G. Simplifying quantum logic using higher-dimensional Hilbert spaces
\textit{Nat. Phys.},\textbf{5}, 134 (2009)

\bibitem{sa-prl-96-083601} Sanaka, K., Resch, K. J. \& Zeilinger, A. Filtering Out Photonic Fock States. \textit{Phys. Rev. Lett.} \textbf{96}, 083601 (2006).
\bibitem{re-prl-98-203602} Resch, K. J., O'Brien, J. L., Weinhold, T. J., Sanaka, K. Lanyon, B. P., Langford, N. K. \& White, A. G. Entanglement Generation by Fock-State Filtration. \textit{Phys. Rev. Lett.} \textbf{98}, 203602 (2007).

\bibitem{ok-sci-323-483} Okamoto, R., O'Brien, J. L, Hofmann, H. F, Nagata, T., Sasaki, K. \& Takeuchi, S.
An Entanglement Filter. \textit{Science} \textbf{323}, 483-485 (2009).

\bibitem{lo-sci-322-563}  Lobino, M., Korystov, D., Kupchak, C., Figueroa, E., Sanders, B. C. \& Lovovsky, A. I. Complete Characterization of Quantum-Optical Processes. \textit{Science} \textbf{322}, 563-566 (2008).
\bibitem{fu-sci-282-706} Furusawa, A., S{\o}rensen, J. L., Braunstein, S. L., Fuchs, C. A., Kimble, H. J. \& Polzik, E. S. Unconditional Quantum Teleportation. \textit{Science} \textbf{282} 706-709 (1998)
\bibitem{pa-sci-317-1890}  Parigi, V., Zavatta, A., Kim, M. \& Bellini, M. Probing Quantum Commutation Rules by Addition and Subtraction of Single Photons to/from a Light Field. \textit{Science} \textbf{317}, 1890-1893 (2007).
\bibitem{br-rmp-77-513} Braunstein, S. L. van Loock, P. Quantum information with continuous variables. \textit{Rev. Mod. Phys.} \textbf{77}, 513 (2005).
\bibitem{ou-sci-312-83} Ourjoumtsev, A., Tualle-Brouri, R., Laurat, J. \& Grangier, P. Generating Optical Schrodinger Kittens for Quantum Information Processing. \textit{Science} \textbf{312}, 83-86 (2006).
\bibitem{me-prl-101-130501} Menicucci, N. C., Flammia, S. T. \& Pfister, O. One-Way Quantum Computing in the Optical Frequency Comb. \textit{Phys. Rev. Lett.} \textbf{101}, 130501 (2008).

\bibitem{32} Mino, S. Recent progress on PLC technologies for large-scale integration. \textit{Optical Fiber Communication and Optoelectronics Conference, 2007 Asia}, 27 (2007).

\bibitem{ke-prl-99-130501} Kieling, K., Ruddolph, T. \& Eisert, J. Percolation, Renormalization, and Quantum Computing with Nondeterministic Gates. \textit{Phys. Rev. Lett.} \textbf{99}, 130501 (2007).

\bibitem{wo-jtsqe-6-69} Wooten, E. L. et. al. A review of lithium niobate modulators for fiber-optic communications systems. \textit{Selected Topics in Quantum Electronics, IEEE Journal of} \textbf{6}, 69-82 (2000).


\bibitem{pe-prl-99-223602} Pezz\'{e}, L., Smerzi, A., Khoury, G., Hodelin, J. F. \& Bouwmeester, D. Phase Detection at the Quantum Limit with Multiphoton Mach-Zehnder Interferometry. \textit{Phys. Rev. Lett.} \textbf{99}, 223602 (2007).


\bibitem{be-pra-73-063824} Berry, D. W. \& Wiseman, H. M. Adaptive phase measurements for narrowband squeezed beams. \textit{Phys. Rev. A.} \textbf{73}, 063824 (2006).





\end{thebibliography}

\vspace{10pt}\begin{large}
{\noindent\textbf{Acknowledgements}}\\
\end{large}
\begin{small}
{\noindent We thank A. Laing, T. Nagata, S. Takeuchi and X. Q. Zhou for helpful discussions. This work was supported by IARPA, EPSRC, QIP IRC  and the Leverhulme Trust.}
\end{small}


\newpage
\begin{large}
\centering
\vspace{12pt}{\noindent\textbf{Supplementary Information}}\\
\vspace{12pt}
\end{large}

We provide here supplementary materials for our Article which details the experimental setup used for the reported measurements, an analysis of the characterization and properties of the resistive heater integrated in the quantum circuit and additional data on the 4-photon integrated quantum metrology experiment.

\vspace{6pt}\noindent\textbf{Calibration of the resistive heater:}
A metal layer patterned on the top of the devices provides resistive elements, metal connections and contact pads that can be used to control locally the temperature of part of the chip.  When a voltage is applied across the contact pads, current flows through $R$ and generates heat which dissipates into the device and locally raises the temperature $T$ of the core and cladding material of the waveguide section directly beneath $R$. A change in the temperature of the waveguide causes a change in refractive index $n$. This induces a change in mode group index of the light confined in the corresponding waveguide section directly beneath \textit{R}, and therefore introduces a phase difference $\phi$ with respect to the unperturbed waveguide. The heat generated inside the resistive elements $R$ dissipates through the depth of the structure to the silicon substrate which acts as a heat sink.

\begin{figure}[b]
\begin{center}
\includegraphics*[width=0.5\textwidth]{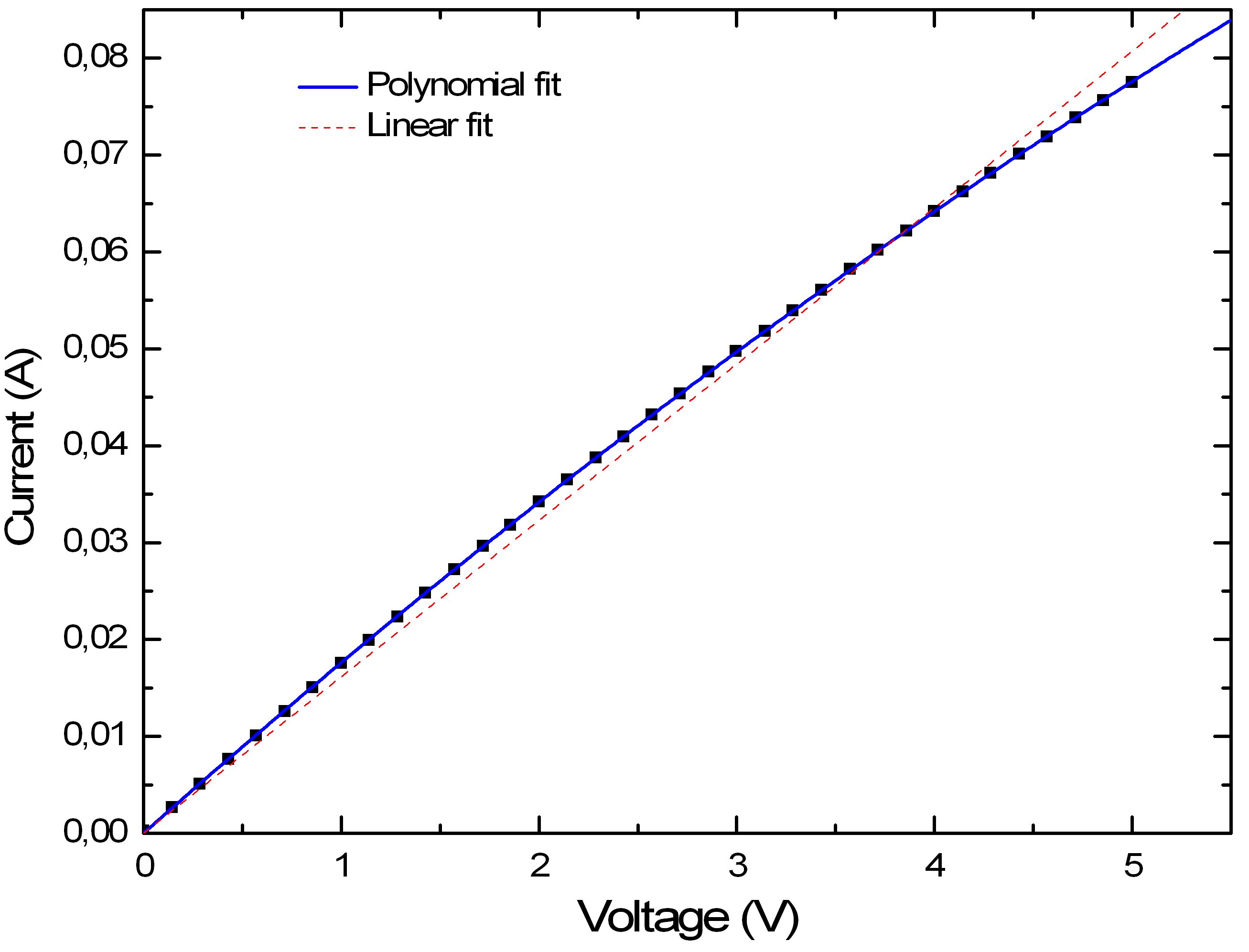}
\vspace{-0.2cm}
\caption{Current-Voltage relation of the resistive heater. To highlight the non-ohmic relation, we show a linear and a polynomial best fit.}
\label{figS5}
\end{center}
\vspace{-0.8cm}
\end{figure}

\begin{figure}[t]
\vspace{-1cm}
\begin{center}
\includegraphics*[width=0.5\textwidth]{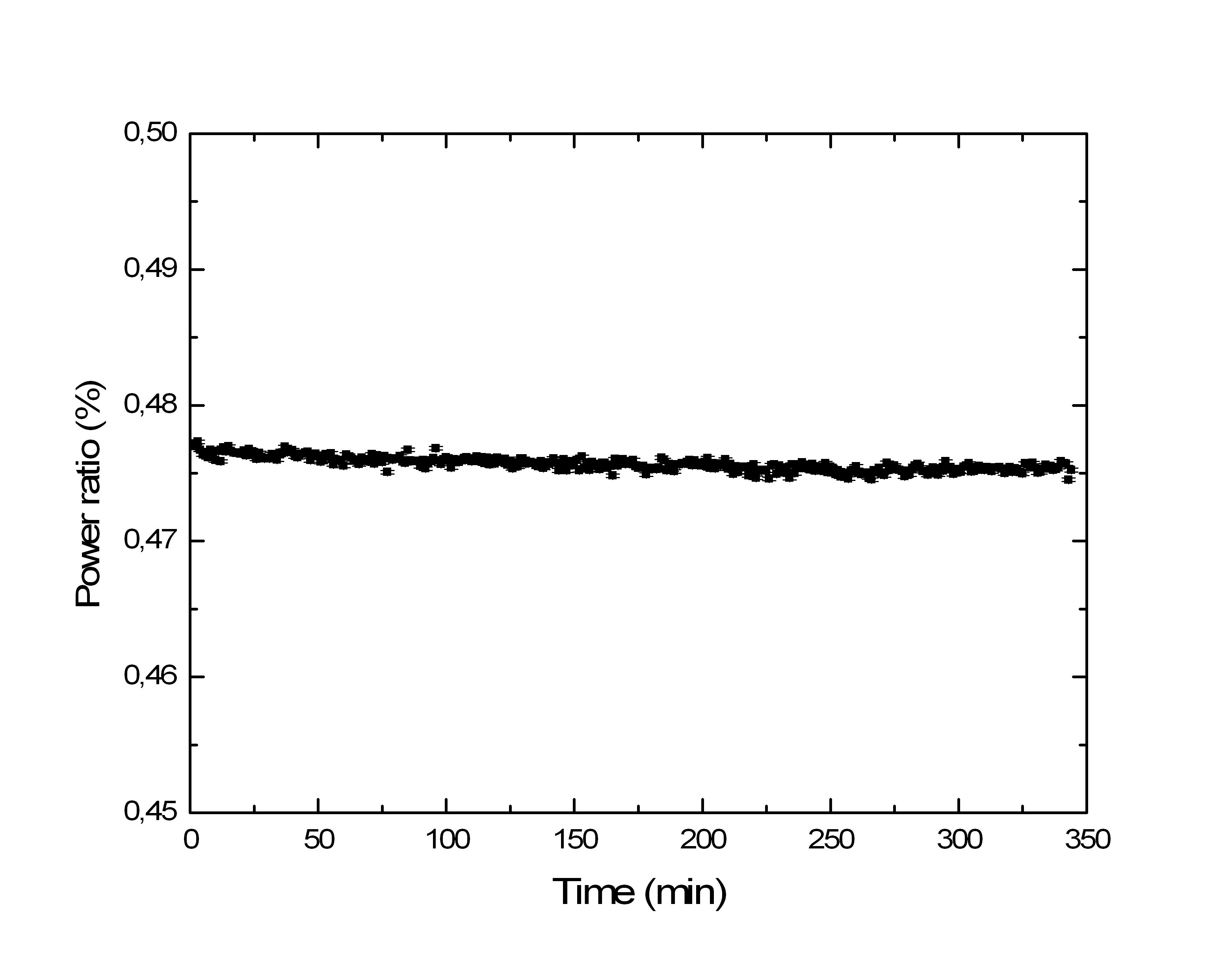}
\vspace{-0.2cm}
\caption{Probability of detecting a photon in mode $h$ when sending a single photon in input $a$ as a function of time. To see the stability of the phase, the probability axis is zoomed in on the range (0.45, 0.5).}
\label{figS1}
\end{center}
\vspace{-0.8cm}
\end{figure}

\begin{table}[b]
\caption{Values of the parameters obtained from the best fit in the calibration of $\phi\left(V\right)$.}
\centering\vspace{6pt}
\begin{tabular}{|c |r @{.} l| l|}
\hline
Parameter & \multicolumn{2}{|c|}{Value} & error\\
\hline
$\alpha$ & -1&887 & 0.006 \\
$\beta$ & 0&157 & 0.005 \\
$\gamma$ & 0&0045 & 0.002 \\
$\delta$ & -0&001 & 0.0002 \\
\hline
\end{tabular}
\end{table}

The voltage-controlled phase inside the interferometer circuit is defined by a nonlinear relation $\phi\left(V\right)$ that depends on the chip fabrication process. For this reason each device has to be calibrated once to relate  the applied voltage to the phase difference introduced in the interferometer. We used a 2-photon experiment to find the $\phi\left(V\right)$ relation modulo $\pi$, since in principle the resulting frequency of  the 2-photon interference pattern is double that of the 1-photon case, allowing a wider sample of an interference fringe. The phase shift ambiguity is later corrected to modulo $2\pi$ by direct comparison to well-known 1-photon ``classical'' interference.  To first approximation the applied phase $\phi$ is proportional to the power dissipated by the resistor. This translates into a quadratic relation between the applied voltage and the phase. To take into account deviations form the ideal case, mainly due to a non-ohmic current-voltage relation as the temperature changes (as shown in Fig. \ref{figS5}), we fixed the form of the $\phi\left(V\right)$ relation as
\begin{eqnarray}
\phi\left(V\right) = \alpha + \beta V^2 + \gamma V^3 + \delta V^4,
\end{eqnarray}
where the $V^3$ term is related to non-ohmic behavior of the resistor, and the $V^4$ term is a combination of non-ohmic relation and the expansion of $\phi\left(V\right)$ in even powers of $V$. The parameters computed from the complete calibration process for $\phi\left(V\right)$ are reported in Table S1, which we use throughout our analysis of the 1-, 2- and 4- photon interference experiments as well as for varying the reflectivity in 2-photon Hong-Ou-Mandel experiments.

An experiment to check the stability of the phase applied in the Mach-Zehnder interferometer was performed using single photons. Fig. \ref{figS1} shows the probability of detecting a photon in mode $h$ when sending single photons in input $a$ as a function of time, a voltage of $1.4V$ was applied across resistive element. The probability remained almost constant for more than six hours. The small deviation is imputed to the different evolution of the coupling from the waveguides $g$ and $h$ to the fiber array that collect the photons at the output of the circuit. However, different evolution of the coupling efficiencies does not lower the quantum mechanical performance of the device.

\begin{figure}[t!]
\vspace{-0.3cm}
\begin{center}
\includegraphics*[width=0.5\textwidth]{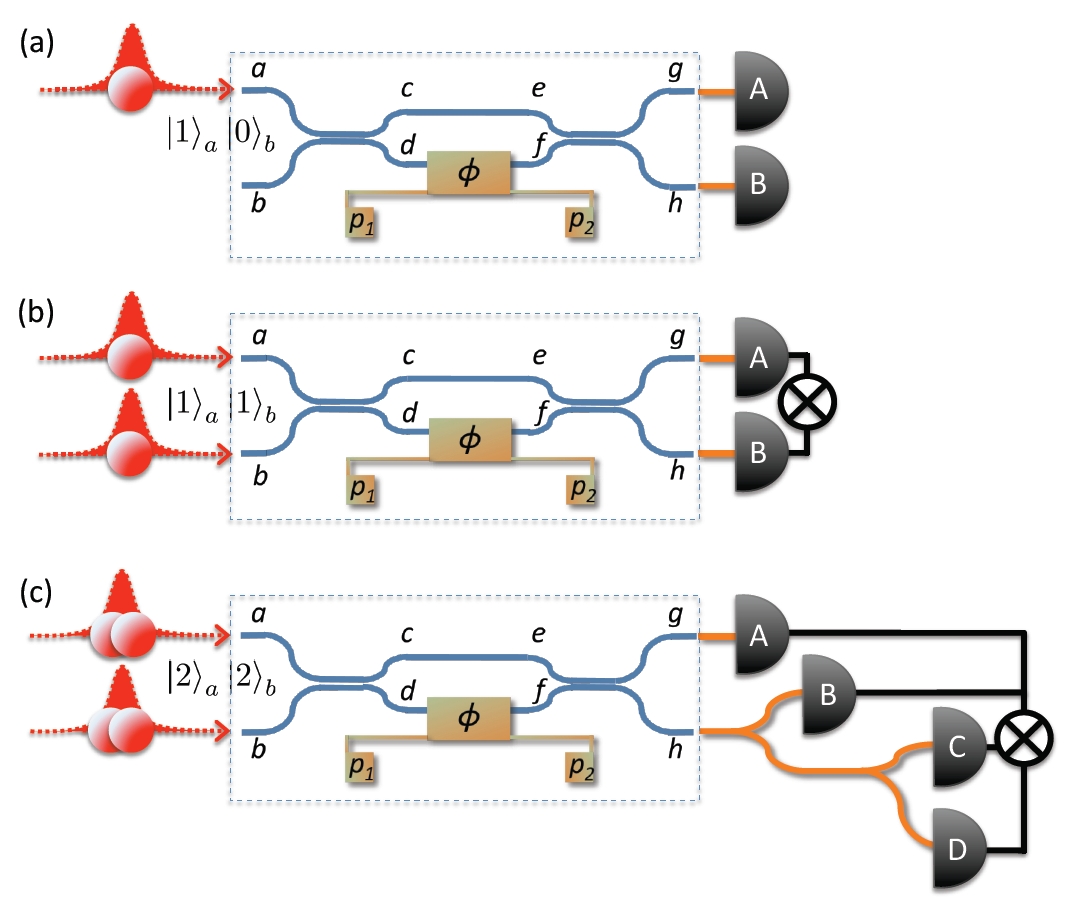}
\vspace{-0.6cm}
\caption{\footnotesize{Experimental schematics for the three interference fringe experiments shown in Fig. 4 of the Article. \textbf{(a)} The 1-photon interference fringe excitation and detection scheme. \textbf{(b)} The 2-photon ``$\lambda/2$'' interference fringe excitation and detection scheme. \textbf{(c)} The 4-photon ``$\lambda/4$'' interference fringe excitation and detection scheme.}}
\label{figS3}
\end{center}
\vspace{-0.3cm}
\end{figure}

\vspace{6pt}\noindent\textbf{Metrology detection schemes:} The 1-, 2- and 4-photon detection schemes for the interference fringe experiments shown in Fig. 4 of the Article are displayed in Fig. \ref{figS3}. The 1-photon interference fringes are observed using one output of the Type-I SPDC coupled to input $a$ of the waveguide circuit; measurement with respect to phase $\phi$ is conducted by coupling each of the outputs $g$ and $h$ to fibre coupled single photon counting modules (SPCMs) labeled A and B and monitoring the respective single photon count rates. The 2-photon interference fringes are observed using degenerate photon pairs coupled from the Type-I SPDC into inputs $a$ and $b$; with respect to $\phi$, the resulting ``$\lambda/2$'' fringes are observed by measuring the 2-fold coincidental photon detection rate across SPCMs A and B. The 4-photon interference fringes are observed using 4-photons emitted in two modes $x$ and $y$ from the pulsed SPDC and coupled into modes $a$ and $b$. The ``$\lambda/4$'' fringe is measured by detecting either the $\ket{3}_g\ket{1}_h$ or $\ket{1}_g\ket{3}_h$ fock states; the latter, for example, is detected non-deterministically by cascading SPCMs B, C and D as shown with two fibre splitters and monitoring the 4-fold coincidental photon detection rate across A, B, C and D. Each fringe is plotted using the phase voltage relation derived from the resistive heater calibration. The same 2-photon setup and detection scheme is used for the the multiple Hong-Ou-Mandel experiments reported in Fig. 5 of the Article, using the $\mu\textrm{m}$ actuator on output $y$ of the SPDC source (shown in Fig. 2 of the Article).

\vspace{6pt}\noindent\textbf{4-photon integrated quantum metrology:}
The 4-photon N00N experiment described in the main article was repeated using a higher pump power from the Ti:Sapphire laser. This was done to confirm with a higher accuracy the reduced de Broglie wavelength by obtaining lower error bars for the experiment. As the power of the pump is increased, the 4-photon production rate increases, but by the same argument, the production of 6-photons start to be non-negligible. This reduces the contrast of the measured fringe visibility, since losses and avalanche detectors that cannot resolve photon number give rise to spurious counts of the $\ket{3}_g\ket{1}_h$ state.

\begin{figure}[b]
\vspace{-0.5cm}
\begin{center}
\includegraphics*[width=0.5\textwidth]{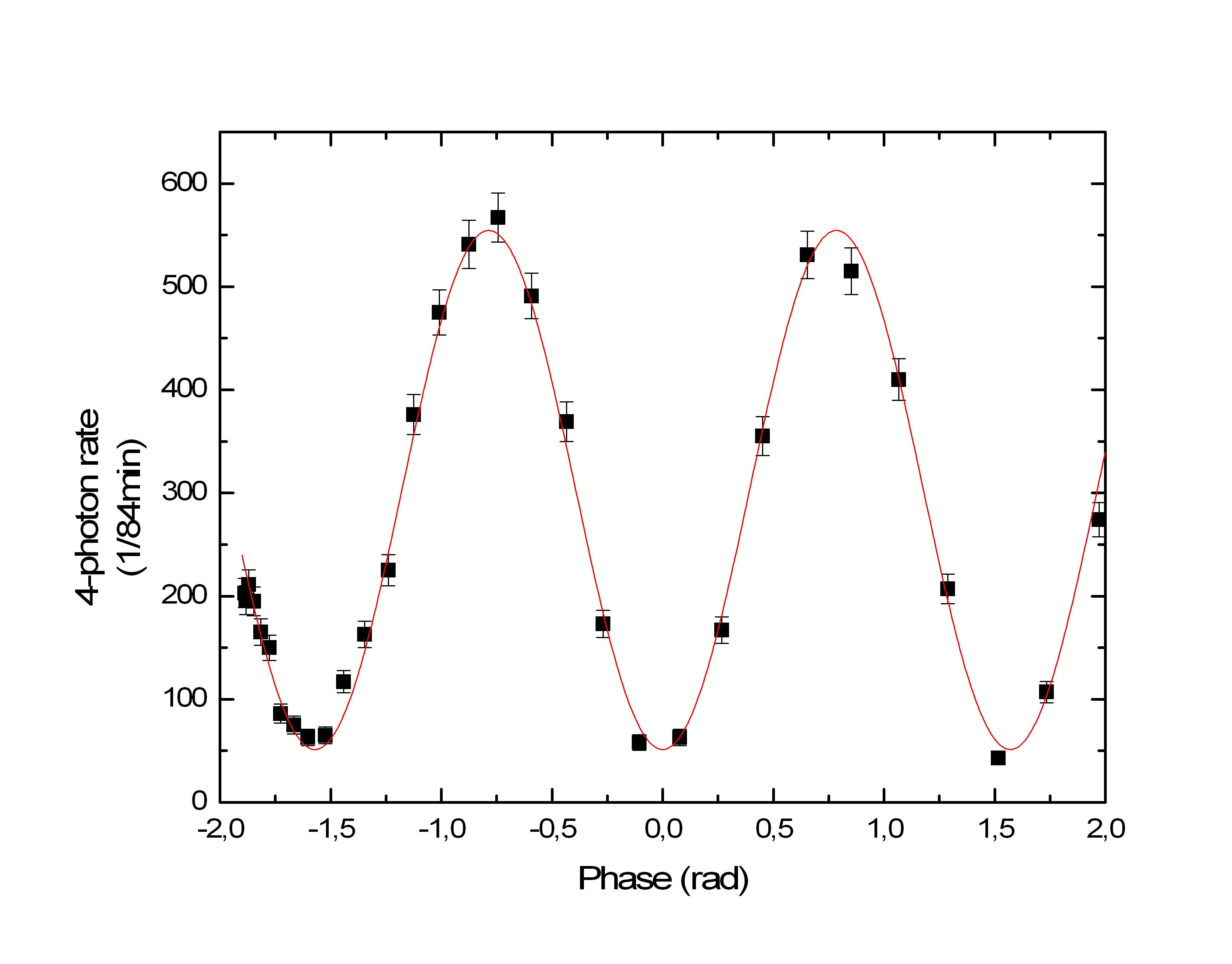}
\vspace{-0.6cm}
\caption{4-photon detection rate of the output state $\ket{3}_g\ket{1}_h$. The experiment was conducted with a high pump power to obtain higher 4-photon count rates.}
\label{4noon}
\end{center}
\vspace{-0.8cm}
\end{figure}

In Fig. \ref{4noon} we show the 4-photon detection rate of the output state $\ket{3}_g\ket{1}_h$ at this higher pump power, while varying the phase $\phi$ of one arm of the interferometer. The contrast of the 4-fold interference fringe is $C=83.1\pm 1.5\%$, which, despite the post-selecting detection scheme and the higher power, is still sufficient to beat the shot noise limit. Note, however, that the reduced visibility is the result of 6 photons being generated at higher pump power.

\end{document}